\documentclass{svjour3}                     % onecolumn (standard format)
\smartqed  % flush right qed marks, e.g. at end of proof
\usepackage{graphicx}
\usepackage{cite}
\journalname{}
\begin{document}

\title{Crossing the Phantom Divide \\
in Extended Dvali--Gabadadze--Porrati Gravity%\thanks{Grants or other notes
}


\author{Koichi Hirano         \and
        Zen Komiya %etc.
}


\institute{K. Hirano \at
              Department of General Education, Ichinoseki National College of Technology, \\
Ichinoseki 021-8511, Japan \\
%              Tel.: +81-191-24-4790\\
%              Fax: \\
              \email{hirano@ichinoseki.ac.jp}           %  \\
           \and
           Z. Komiya \at
              Department of Physics, Tokyo University of Science, Tokyo 162-8601, Japan \\
Shinjuku College of Information Technology, Tokyo 164-0001, Japan
}

\date{Received: date / Accepted: date}

\maketitle

\begin{abstract}
We propose a phantom crossing Dvali--Gabadadze--Porrati (DGP) model. In our model, the effective equation of state of the DGP gravity crosses the phantom divide line. We demonstrate crossing of the phantom divide does not occur within the framework of the original DGP model or the DGP model developed by Dvali and Turner. By extending their model, we construct a model that realizes crossing of the phantom divide. We find that the smaller the value of the new introduced parameter $\beta$ is, the older epoch crossing of the phantom divide occurs in. Our model can account for late-time acceleration of the universe without dark energy. We investigate and show the property of Phantom crossing DGP model.
%\PACS{95.36.+x \and 98.80.Es}
\end{abstract}

\section{Introduction}
Late time accelerated expansion of the universe was indicated by measurements of distant Type Ia supernovae (SNe Ia) \cite{rie1998, per1999, kno2003, rie2004, rie2007, ast2006, mik2007, woo2007, fri2008}.
 This was confirmed by observations of cosmic microwave background (CMB) anisotropies by the Wilkinson Microwave Anisotropy Probe (WMAP) \cite{spe2007, kom2009}, and the large-scale structure in the distribution of galaxies observed in the Sloan Digital Sky Survey (SDSS) \cite{teg2004a, teg2004b}.
\par
It is not possible to account for this phenomenon within the framework of general relativity containing only matter. Therefore, a number of models containing "dark energy" have been proposed as the mechanism for the acceleration. There are currently many dark energy models, including cosmological constant, scalar field, quintessence, and phantom models \cite{rat1988, pee1988, cal2002, cal2003, hir2008, har2008, hir2006, kom2005, kom2006}.
However, dark energy, the nature of which remains unknown, has not been detected yet. The cosmological constant, which is the standard candidate for dark energy, cannot be explained by current particle physics due to its very small value, and it is plagued with fine-tuning problems and the coincidence problem.
\par
An alternative method for explaining the current accelerated expansion of the universe is to extend general relativity to more general theories on cosmological scales. Instead of adding an exotic component such as a cosmological constant to the right-hand side (i.e., the energy-momentum tensor) of Einstein's field equation, the left-hand side (i.e., the Einstein tensor, which is represented by pure geometry) can be modified. Typical models based on this modified gravity approach are $f(R)$ models \cite{noj2007, noj2008a, noj2008b} and the Dvali--Gabadadze--Porrati (DGP) model \cite{dva2000, def2001, def2002} (for reviews, see \cite{koy2007}).
\par
In $f(R)$ models, the scalar curvature $R$ in the standard Einstein--Hilbert gravitational Lagrangian is replaced by a general function $f(R)$. By adopting appropriate function phenomenologically, $f(R)$ models can account for late-time acceleration without postulating dark energy.
\par
The DGP model is an extra dimension scenario. In this model, the universe is considered to be a brane; i.e., a four-dimensional (4D) hypersurface, embedded in a five-dimensional (5D) Minkowski bulk. On large scales, the late-time acceleration is driven by leakage of gravity from the 4D brane into 5D spacetime. Naturally, there is no need to introduce dark energy. On small scales, gravity is bound to the 4D brane and general relativity is recovered to a good approximation.
\par
According to various recent observational data including that of Type Ia supernovae \cite{ala2004, nes2007, wuy2006, jas2006}, it is possible that the effective equation of state parameter $w_{\rm eff}$, which is the ratio of the effective pressure $p_{\rm eff}$ to the effective energy density $\rho_{\rm eff}$, evolves from being larger than $-1$ (non-phantom phase) to being less than $-1$ (phantom phase \cite{cal2002, noj2003}); namely, it has currently crossed $-1$ (the phantom divide).
\par
$f(R)$ models that realize the crossing of the phantom divide have been studied \cite{bam2009b, bam2009a}. On the other hand, in the original DGP model \cite{dva2000, def2001, def2002} and a phenomenological extension of the DGP model described by the modified Friedmann equation proposed by Dvali and Turner \cite{dva2003}, the effective equation of state parameter never crosses the $w_{\rm eff}$ = $-1$ line.\footnote{Some models realize phantom crossing by adding an exotic component such as the scalar field to the DGP model \cite{noz2008, noz2009}. In these models, the equation of state of the additional component crosses the phantom divide, but the effective equation of state of DGP gravity itself does not.}
\par
In this paper, we develop the "Phantom Crossing DGP model" by further extending the modified Friedmann equation by Dvali and Turner \cite{dva2003}. In our model, the effective equation of state parameter of DGP gravity crosses the phantom divide line, as indicated by recent observations.
\par
This paper is organized as follows. In the next section, we summarize the original DGP model, and check the behavior of the effective equation of state. In Section \ref{sec:3}, we describe the modified Friedmann equation by Dvali and Turner \cite{dva2003}, and we also demonstrate that the effective equation of state does not cross the $w_{\rm eff}$ = $-1$ line in this framework. In Section \ref{sec:4}, we construct "the Phantom Crossing DGP model" by extending the modified Friedmann equation proposed by Dvali and Turner. We show that the effective equation of state parameter of our model crosses the phantom divide line, and investigate the properties of our model. Finally, a summary is given in Section \ref{sec:5}.
\section{The DGP model}
The DGP model \cite{dva2000} assumes that we live on a 4D brane embedded in a 5D Minkowski bulk. Matter is trapped on the 4D brane and only gravity experiences the 5D bulk. 
\par
The action is
\begin{eqnarray}
S & = & \frac{1}{16\pi}M_{(5)}^3\int_{bulk}{d^5x\sqrt{-g_{(5)}}R_{(5)}} \nonumber \\
  &   & + \frac{1}{16\pi}M_{(4)}^2\int_{brane}{d^4x\sqrt{-g_{(4)}}(R_{(4)}+L_m)},
\end{eqnarray}
where the subscripts (4) and (5) denote quantities on the brane and in the bulk, respectively. $M_{(5)}$ ($M_{(4)}$) is the 5D (4D) Planck mass, and $L_m$ represents the matter Lagrangian confined on the brane. The transition from 4D gravity to 5D gravity is governed by a crossover scale $r_c$.
\begin{equation}
r_c = \frac{M_{(4)}^2}{2M_{(5)}^3}.
\end{equation}
On scales larger than $r_c$, gravity appears 5D. On scales smaller than $r_c$, gravity is effectively bound to the brane and 4D Newtonian dynamics is recovered to a good approximation. $r_c$ is the single parameter in this model.
\par
Assuming spatial homogeneity and isotropy, a Friedmann-like equation on the brane is obtained as \cite{def2001, def2002}
\begin{equation}
H^2 = \frac{8\pi G}{3}\rho+\epsilon\frac{H}{r_c} \label{dgp_fri},
\end{equation}
where $\rho$ is the total cosmic fluid energy density on the brane. $\epsilon = \pm 1$ represents the two branches of the DGP model. The solution with $\epsilon = +1$ is known as the self-accelerating branch. In this branch, the expansion of the universe accelerates even without dark energy because the Hubble parameter approaches a constant, $H = 1/r_c$, at late times. On the other hand, $\epsilon = -1$ corresponds to the normal branch. This branch cannot undergo acceleration without an additional dark energy component. Hence in what follows we consider the self-accelerating branch ($\epsilon = +1$) only.
\par
For the second term on the right-hand side of Eq. (\ref{dgp_fri}), which represents the effect of DGP gravity, the effective energy density is
\begin{equation}
\rho_{r_c} = \frac{3}{8\pi G}\frac{H}{r_c}, \label{rho_eff}
\end{equation}
and the effective pressure is
\begin{equation}
P_{r_c} =  -\frac{1}{8\pi G}\left(\frac{\dot{H}}{r_cH}+3\frac{H}{r_c}\right), \label{p_eff}
\end{equation}
where $\dot{H}=dH/dt$, the differential of the Hubble parameter with respect to the cosmological time $t$.
Using Eqs. (\ref{rho_eff}) and (\ref{p_eff}), the effective equation of state parameter of DGP gravity is given by
\begin{equation}
w_{r_c} = \frac{P_{r_c}}{\rho_{r_c}}.
\end{equation}
\begin{figure}[h!]
\includegraphics{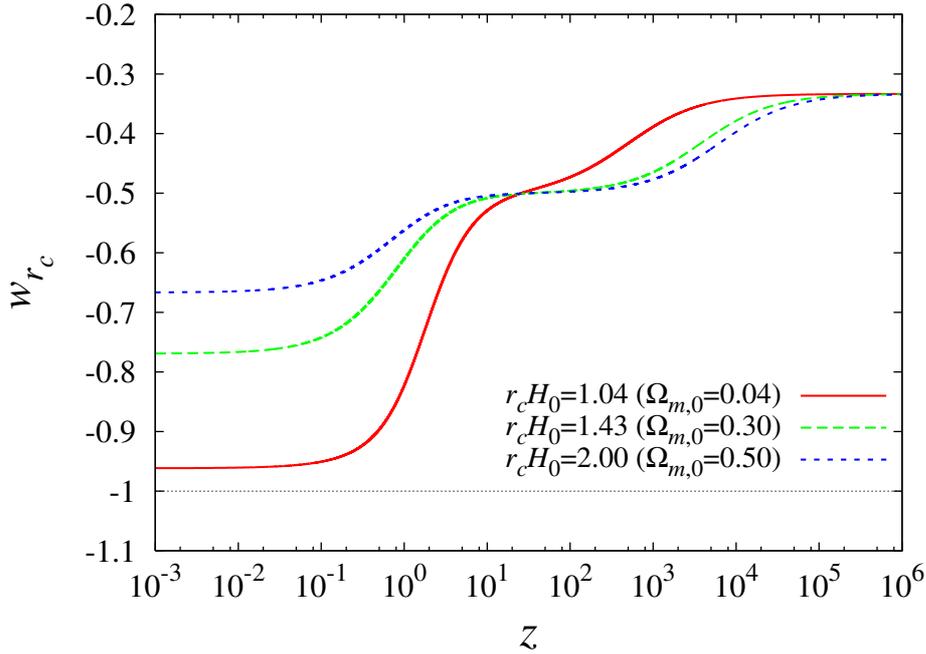}
\caption{Effective equation of state of the DGP model $w_{r_c}$ vs. redshift $z$. The red (solid), green (dashed), blue (dotted) lines represent the cases for $r_cH_0 = 1.04, 1.43$, and $2.00$, respectively (corresponding to $\Omega_{m,0} = 0.04, 0.30$, and $0.50$, respectively).
\label{fig:dgp_wrc}}
\end{figure}
\par
Fig. \ref{fig:dgp_wrc} shows the behavior of the effective equation of state of DGP gravity $w_{r_c}$ versus the redshift $z$ for $r_cH_0 = 1.04, 1.43$, and $2.00$. Assuming that the total cosmic fluid energy density $\rho$ of Eq. (\ref{dgp_fri}) contains matter and radiation, from Eq. (\ref{rch0omegam}), these values of $r_cH_0$ correspond to $\Omega_{m,0} = 0.04, 0.30$, and $0.50$, respectively.
\begin{equation}
\Omega_{m,0} = 1 - \Omega_{r,0} - (r_cH_0)^{-1}, \label{rch0omegam}
\end{equation}
where $\Omega_m$ is the normalized energy density of matter and $\Omega_r$ is the radiation on the brane; i.e., $\Omega_m = (8\pi G/3H^2)\rho_m$ and $\Omega_r = (8\pi G/3H^2)\rho_r$. ($\rho_m\propto a^{-3}$, $\rho_r\propto a^{-4}$). The subscripts $0$ designate the present value.
\par
The effective equation of state of DGP $w_{r_c}$ can also be exactly expressed in terms of the energy densities of matter and radiation, $\Omega_m$ and $\Omega_r$ \cite{lue2004, lue2006}.
\begin{equation}
w_{r_c} = - \frac{1}{1+\Omega_m+\Omega_r}.
\end{equation}
In realistic ranges of the energy density, $\Omega_m > 0$ and $\Omega_r \ge 0$, the value of the effective equation of state cannot be less than or equal to $-1$. That is, the effective equation of state never crosses the phantom divide line in the original DGP model.
\section{DGP model extended by Dvali and Turner \label{sec:3}}
Dvali and Turner \cite{dva2003} phenomenologically extended the Friedmann-like equation (Eq. (\ref{dgp_fri})) of the DGP model. This model interpolates between the original DGP model and the pure $\Lambda$CDM model with an additional parameter $\alpha$. The modified Friedmann-like equation is \cite{dva2003}
\begin{equation}
H^2 = \frac{8\pi G}{3}\rho+\frac{H^{\alpha}}{{r_c}^{2-\alpha}}. \label{dt_fri}
\end{equation}
For $\alpha = 1$, this agrees with the original DGP Friedmann-like equation, while $\alpha = 0$ leads to an expansion history identical to that of $\Lambda$CDM cosmology.
\par
Differentiating both sides of Eq. (\ref{dt_fri}) with respect to the cosmological time $t$, we obtain the following differential equation.
\begin{equation}
2\dot{H}=-8\pi G(\rho+P)+\frac{\alpha\dot{H}}{(r_cH)^{2-\alpha}}, \label{hdot2}
\end{equation}
where a dot indicates the derivative respect to the cosmological time. The quantity $P$ is the total cosmic fluid pressure on the brane.
\par
For the second term on the right-hand side of Eq. (\ref{dt_fri}), which represents the effect of DGP gravity, the effective energy density is
\begin{equation}
\rho_{\alpha}=\frac{3}{8\pi G}\frac{H^{\alpha}}{{r_c}^{2-\alpha}}, \label{rho_dt}
\end{equation}
and from Eq. (\ref{hdot2}) the effective pressure is
\begin{equation}
P_{\alpha}=-\frac{1}{8\pi G}\left[\frac{\alpha\dot{H}}{(r_cH)^{2-\alpha}}+3\frac{H^{\alpha}}{{r_c}^{2-\alpha}}\right]. \label{p_dt}
\end{equation}
From Eqs. (\ref{rho_dt}) and (\ref{p_dt}), the effective equation of state parameter of the DGP model extended by Dvali and Turner is given by
\begin{equation}
w_{\alpha} = \frac{P_{\alpha}}{\rho_{\alpha}}. \label{w_dt}
\end{equation}
\par
\begin{figure}[h!]
\includegraphics{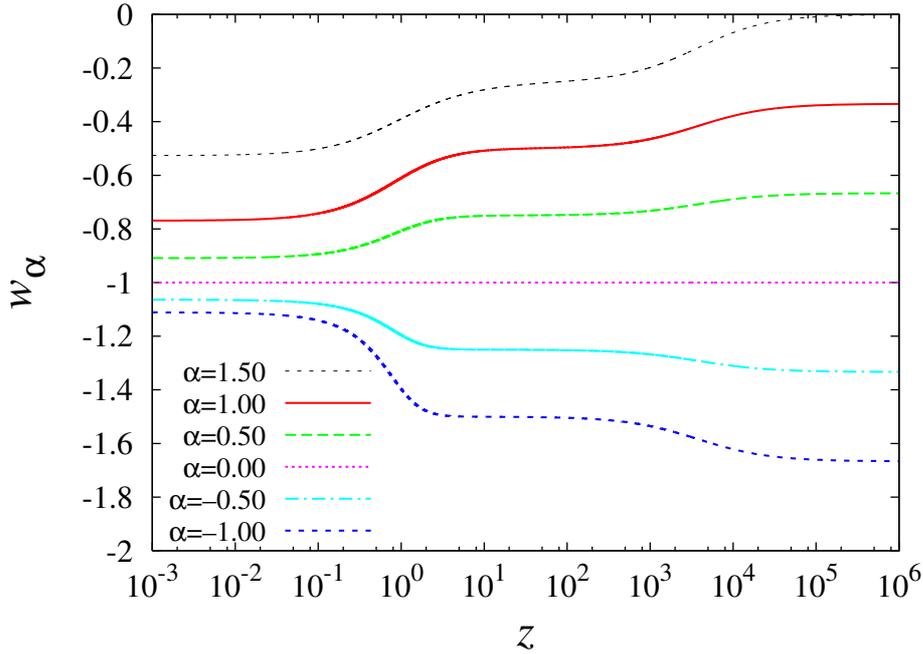}
\caption{Effective equation of state $w_{\alpha}$, vs. redshift $z$ of the DGP model extended by Dvali and Turner for $\alpha = 1.50, 1.00, 0.50, 0.00, -0.50$, and $-1.00$ (top to bottom) assuming $\Omega_{m,0} = 0.30$.
\label{fig:dt_wa}} 
\end{figure}

Fig. \ref{fig:dt_wa} shows a plot of the behavior of the effective equation of state of the DGP model by Dvali and Turner $w_\alpha$ versus the redshift $z$ for $\alpha = 1.50, 1.00, 0.50, 0.00, -0.50$, and $-1.00$ (assuming $\Omega_{m,0} = 0.30$). 
\par
In general, for equation of state $w$, the energy density $\rho$ varies as $a^{-3(1+w)}$. This leads to the following proportional relation.
\begin{equation}
\rho_{\alpha} \propto a^{-3(1+w_{\alpha})}. \label{rad1}
\end{equation}
At the same time, From Eq. (\ref{rho_dt}), we find $\rho_{\alpha} \propto H^{\alpha}$. In the radiation-dominated epoch, from the proportional relation on Hubble parameter $H \propto a^{-2}$, we obtain the following relation.
\begin{equation}
\rho_{\alpha} \propto a^{-2\alpha}. \label{rad2}
\end{equation}
As compared the right-hand side of Eq. (\ref{rad1}) to that of Eq. (\ref{rad2}), during the earlier radiation-dominated epoch ($z\gg10^4$), the effective equation of state can also be represented 
% by the following equation 
with $\alpha$ \cite{dva2003}.
\begin{equation}
w_\alpha = -1 + \frac{2\alpha}{3}. \label{wa_rad}
\end{equation}
In the same way, during the matter-dominated epoch ($10^2$\hspace{0.3em}\raisebox{0.4ex}{$>$}\hspace{-0.75em}\raisebox{-.7ex}{$\sim$}\hspace{0.3em}$z$$\gg$$1$), from the proportional relation on Hubble parameter $H \propto a^{-\frac{3}{2}}$,
\begin{equation}
w_\alpha = -1 + \frac{\alpha}{2}.
\end{equation}
At the present time, $w_\alpha$ is a stationary value close to $-1$.
\par
From these results, in the case of $\alpha < 1.0$, the effective equation of state $w_\alpha < -\frac{1}{3}$ in all era, even in the radiation-dominated epoch. That is to say, the component of the DGP gravity works as the driving force of the accelerated expansion of the universe in all epochs. On the other hand, for $\alpha \hspace{0.3em}\raisebox{0.4ex}{$>$}\hspace{-0.75em}\raisebox{-.7ex}{=}\hspace{0.3em} 1.0$, there is era when the effective equation of state becomes $w_\alpha \hspace{0.3em}\raisebox{0.4ex}{$>$}\hspace{-0.75em}\raisebox{-.7ex}{=}\hspace{0.3em} -\frac{1}{3}$. That is, the DGP gravity does not drive the accelerated expansion in all epochs.
\par
The case of $\alpha = 1.0$ corresponds to the original DGP model described in the previous section. Thus, in the original DGP model, the effective equation of state $w_{r_c} = -\frac{1}{3}$ in the radiation-dominated epoch. And after the radiation-dominated epoch, becomes $w_{r_c} < -\frac{1}{3}$. In other words, the DGP gravity acts as the driving force of the accelerated expansion just after the radiation-dominated epoch.
\par
However, when $\alpha$ is positive, the effective equation of state $w_\alpha$ will exceed $-1$ at all times. For negative $\alpha$, $w_\alpha$ is always less than $-1$. In the case of $\alpha = 0$, $w_\alpha$ is $-1$ constantly. Based on this analysis, crossing of the phantom divide does not occur in the DGP model extended by Dvali and Turner.
\section{Phantom crossing DGP model \label{sec:4}}
We propose the "Phantom Crossing DGP model" that extends the modified Friedmann equation (Eq. (\ref{dt_fri})) proposed by Dvali and Turner. Our model can realize crossing of the phantom divide line for the effective equation of state of the DGP gravity.
\par
As mentioned in the previous section, the effective equation of state parameter of the DGP model by Dvali and Turner $w_\alpha$, takes the value of over $-1$ for positive $\alpha$, and it is less than $-1$ for negative $\alpha$. When $\alpha = 0$, $w_\alpha$ becomes $-1$. On the basis of these results, we consider a model in which $\alpha$ varies being positive to being negative. To keep the model as simple as possible, we make the following assumption, 
\begin{equation}
\alpha = \beta - a, \label{beta}
\end{equation}
where $a$ is the scale factor (normalized such that the present day value is unity). The quantity $\beta$ is a constant parameter. In the period when the scale factor $a$ is less than the parameter $\beta$ ($\alpha > 0$), the effective equation of state exceeds $-1$. At the point when the scale factor $a$ equals $\beta$, ($\alpha = 0$), the equation of state's value will be $-1$. In the period when the scale factor $a$ exceeds the parameter $\beta$ ($\alpha < 0$), the equation of state will be less than $-1$. In this way, crossing of the phantom divide is realized in our model.
\par
Replacing $\alpha$ by $\beta - a$ in Eq. (\ref{dt_fri}), the Friedmann-like equation in our model is given by
\begin{equation}
H^2 = \frac{8\pi G}{3}\rho+\frac{H^{\beta-a}}{{r_c}^{2-(\beta-a)}}. \label{hirano_fri}
\end{equation}
Differentiating both sides of Eq. (\ref{hirano_fri}) with respect to the cosmological time $t$, the following differential equation is obtained.
\begin{equation}
2\dot{H}=-8\pi G(\rho+P)+\frac{(\beta-a)\dot{H}-\dot{a}H\ln{(r_cH)}}{(r_cH)^{2-(\beta-a)}}. \label{hirano_hdot2}
\end{equation}
For the second term on the right-hand side of Eq. (\ref{hirano_fri}) representing the effect of DGP gravity, the effective energy density is
\begin{equation}
\rho_{\beta}=\frac{3}{8\pi G}\frac{H^{\beta - a}}{{r_c}^{2-(\beta - a)}}, \label{rho_hirano}
\end{equation}
and from Eq. (\ref{hirano_hdot2}), the effective pressure is
\begin{equation}
\hspace*{-7mm} P_{\beta}=-\frac{1}{8\pi G}\left[\frac{(\beta-a)\dot{H}-\dot{a}H\ln{(r_cH)}}{(r_cH)^{2-(\beta-a)}}+3\frac{H^{\beta-a}}{{r_c}^{2-(\beta-a)}}\right]. \label{p_hirano}
\end{equation}
Using Eqs. (\ref{rho_hirano}) and (\ref{p_hirano}), the effective equation of state of our model is given by
\begin{equation}
w_{\beta} = \frac{P_{\beta}}{\rho_{\beta}}. \label{w_hirano}
\end{equation}
\begin{figure}[t]
\includegraphics[width=118mm]{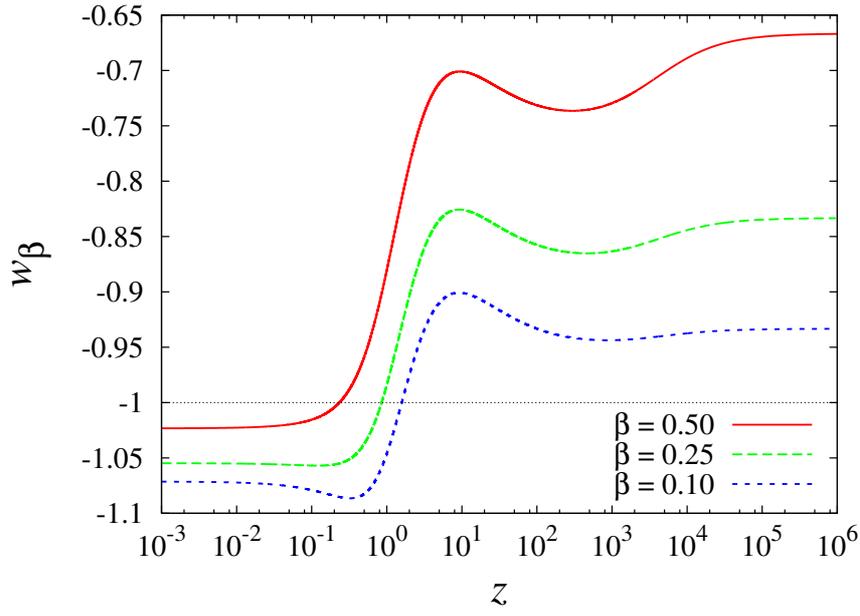}
\caption{Effective equation of state of our model $w_{\beta}$, vs. redshift $z$. The red (solid), green (dashed), blue (dotted) lines represent the cases of $\beta = 0.50, 0.25$, and $0.10$, respectively (assuming $\Omega_{m,0} = 0.30$).
\label{fig:dgp_hirano}}
\end{figure}
\begin{figure}[bh!]
\includegraphics[width=118mm]{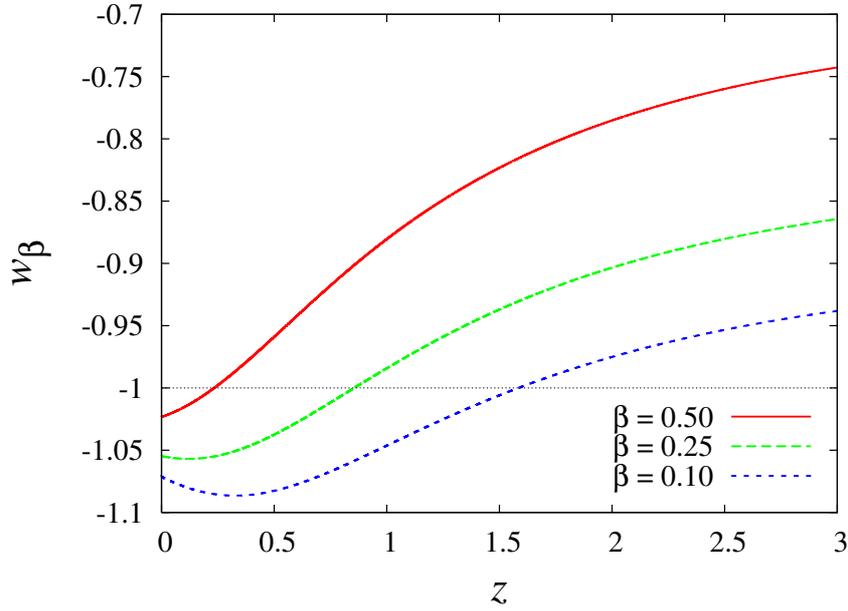}
\caption{Detail of the behavior of $w_{\beta}$ depicted in Fig. \ref{fig:dgp_hirano} near recent epochs.
\label{fig:dgp_hirano_kakudai}} 
\end{figure}
\par
Fig. \ref{fig:dgp_hirano} shows a plot of the effective equation of state of our model $w_\beta$ versus the redshift $z$ (see also Fig. \ref{fig:dgp_hirano_kakudai} which shows an enlarged view of this diagram). Our model is an extension of the DGP model and realizes crossing of the phantom divide. The effective equation of state $w_\beta$ of models for $\beta = 0.50, 0.25,$ and $0.10$ (assuming $\Omega_{m,0} = 0.30$) crosses the phantom divide line when the redshift $z \sim 0.2, 0.8,$ and $1.6$, respectively.
\par
We find that the smaller the parameter $\beta$ is, the older epoch crossing of the phantom divide occurs in. $\beta$ is not necessarily equal to the scale factor at the time of crossing the phantom divide, even though Eq. (\ref{beta}) is assumed. In the Eq. (\ref{hirano_fri}), the value of $\beta - a$ that is the power index of $H$ varies with respect to time. As the power index of the differential equation changes with time, furthermore, in parallel, the differential equation is solved with respect to time. Hence, the time lag occurs, the scale factor at the time of crossing the phantom divide is more than the value of $\beta$.
\par
In a way similar to the derivation of Eq. (\ref{wa_rad}), we represent the effective equation of state of Phantom Crossing DGP model with $\beta$. In the radiation-dominated epoch, the scale factor $a$ is taken to be $0$ in comparison with the value of $\beta$. Therefore, as $\alpha\approx\beta$ in Eq. (\ref{beta}), during the radiation-dominated epoch ($z\gg10^4$), the effective equation of state is approximately
\begin{equation}
w_\beta \approx -1 + \frac{2\beta}{3}.
\end{equation}
That is, in the case of $\beta < 1.0$, the effective equation of state $w_\beta < -\frac{1}{3}$ in all era, including the radiation-dominated epoch. On the other hand, for $\beta \hspace{0.3em}\raisebox{0.4ex}{$>$}\hspace{-0.75em}\raisebox{-.7ex}{=}\hspace{0.3em} 1.0$, there is era when the effective equation of state becomes $w_\beta \hspace{0.3em}\raisebox{0.4ex}{$>$}\hspace{-0.75em}\raisebox{-.7ex}{=}\hspace{0.3em} -\frac{1}{3}$.
\par
\begin{figure}[b]
\includegraphics{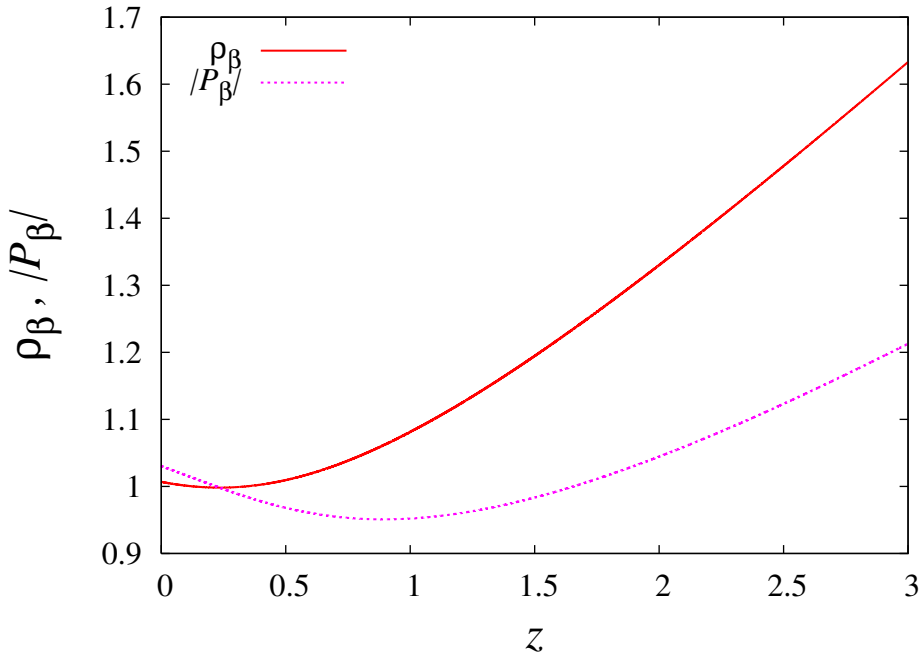}
\caption{The red (solid) and pink (dotted) lines respectively represent the effective energy density $\rho_\beta$ and absolute value of the effective pressure $|P_\beta|$ (note that $P_\beta < 0$) of our model vs. redshift $z$, for ($\beta, \Omega_{m,0}$) = ($0.50, 0.30$).
\label{fig:rhop_hirano}}
\end{figure}
\par
Fig. \ref{fig:rhop_hirano} shows the effective energy density $\rho_\beta$ and absolute value of the effective pressure $|P_\beta|$ (note that $P_\beta < 0$) of our model for ($\beta, \Omega_{m,0}$) = ($0.50, 0.30$) versus the redshift $z$, normalized such that the effective energy density is unity at the time of phantom crossing. It shows that the absolute value of the effective pressure $|P_\beta|$ exceeds the effective energy density $\rho_\beta$ at the time of crossing of the phantom divide.
\par
The recent observational data for Type Ia supernovae \cite{rie2007} show that crossing of the phantom divide line occurs at a redshift $z \sim 0.2$ \cite{ala2004, nes2007, wuy2006}.\footnote{This is a model-dependent value. We will investigate in detail the allowed parameter region based on recent observational data in future work.} In our model, for $\beta = 0.50$ (when $\Omega_{m,0} = 0.30$), crossing of the phantom divide occurs at $z \sim 0.2$.
\\
\par
In a proposed model in which the phantom divide is crossed at $z \sim 0.2$, ($\beta, \Omega_{m,0}$) = ($0.50, 0.30$), we investigate and show the property of Phantom crossing DGP model.
\begin{figure}[h!]
\includegraphics{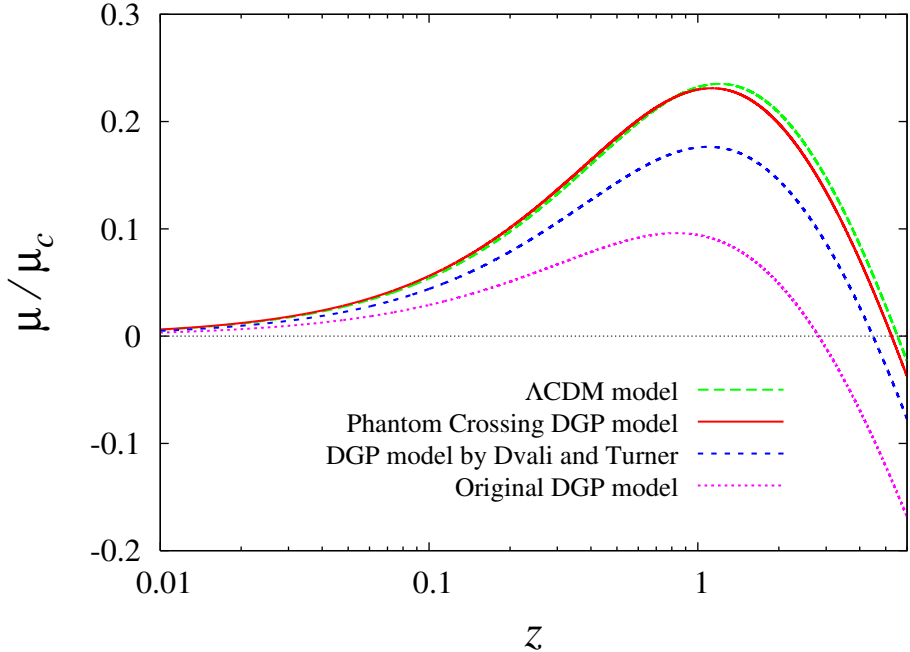}
\caption{The distance modulus $\mu$ relative to that of a constant expansion cosmology $\mu_{c}$, vs. the redshift $z$. Models and parameters are (from top to bottom): (1) $\Lambda$CDM model, $\Omega_{m,0}$ = 0.30; (2) Phantom Crossing DGP model, $\beta$ = 0.50, $\Omega_{m,0}$ = 0.30; (3) DGP model by Dvali and Turner, $\alpha$ = 0.50, $\Omega_{m,0}$ = 0.30; (4) Original DGP model, $\Omega_{m,0}$ = 0.30.
\label{fig:snia}}
\end{figure}
\begin{figure}[h!]
\includegraphics{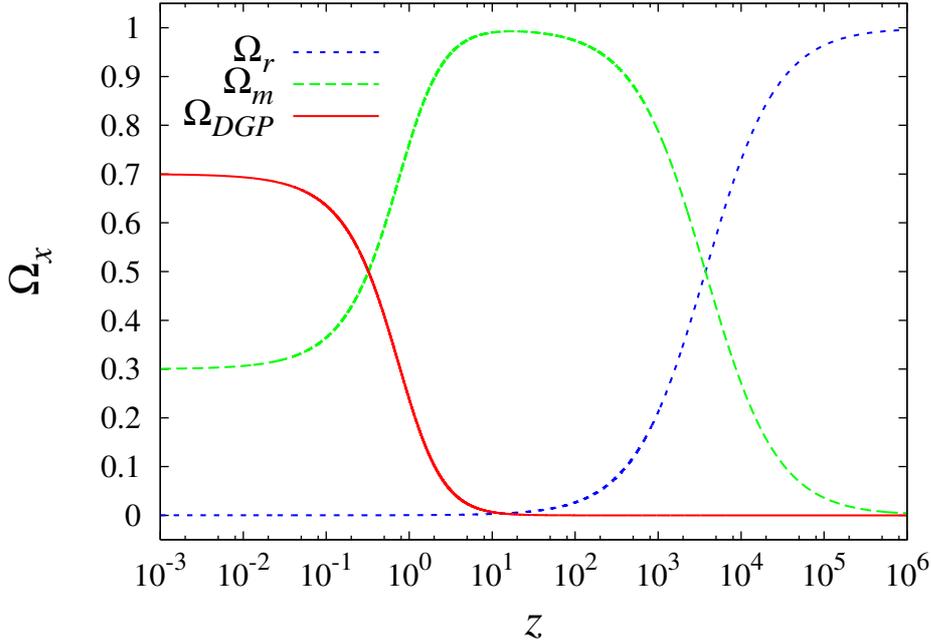}
\caption{The normalized energy density of radiation $\Omega_r$, matter $\Omega_m$, and DGP gravity $\Omega_{DGP}$, vs. the redshift $z$ in the Phantom Crossing DGP model with the proposed parameter ($\beta, \Omega_{m,0}$) = ($0.50, 0.30$).
\label{fig:omega}}
\end{figure}
\par
Fig. \ref{fig:snia} shows the distance modulus $\mu$ relative to that of a constant expansion cosmology $\mu_{c}$, versus the redshift $z$. That is, when $\mu /\mu_{c}$ is positive, cosmic expansion is accelerating. The distance modulus is defined by
\begin{equation}
\mu(z)=5\log_{10}{D_L} + 42.38 - 5\log_{10}{h},
\end{equation}
where $D_L$ is the Hubble free luminosity distance given by
\begin{equation}
D_L=(1+z)\int^z_0\frac{H_0}{H(z^{\prime})}dz^{\prime},
\end{equation}
$h$ being the Hubble constant $H_0$ in units of $100~{\rm km~s^{-1}~Mpc^{-1}}$. We adopt $h = 0.72$ \cite{fre2001}. In Fig. \ref{fig:snia}, Models and parameters are (from top to bottom): (1) $\Lambda$CDM model, $\Omega_{m,0}$ = 0.30; (2) Phantom Crossing DGP model, $\beta$ = 0.50, $\Omega_{m,0}$ = 0.30; (3) DGP model by Dvali and Turner, $\alpha$ = 0.50, $\Omega_{m,0}$ = 0.30; (4) Original DGP model, $\Omega_{m,0}$ = 0.30. Phantom Crossing DGP model can realize late-time acceleration of the universe very similar to that for $\Lambda$CDM model, without dark energy.
\par
Fig. \ref{fig:omega} shows the normalized energy density of radiation $\Omega_r$, matter $\Omega_m$, and DGP gravity $\Omega_{DGP}$ versus the redshift $z$ in the Phantom Crossing DGP model with the proposed parameter ($\beta, \Omega_{m,0}$) = ($0.50, 0.30$). Where, $\Omega_{DGP} = (8\pi G/3H^2)\rho_{\beta}$. $\rho_{\beta}$ is the effective energy density of DGP gravity defined by Eq. (\ref{rho_hirano}). We find that the universe is DGP gravity-dominated near recent epochs. Therefore, In the Phantom Crossing DGP model, the late-time acceleration is driven by the effect of DGP gravity.
\begin{figure}[h!]
\includegraphics{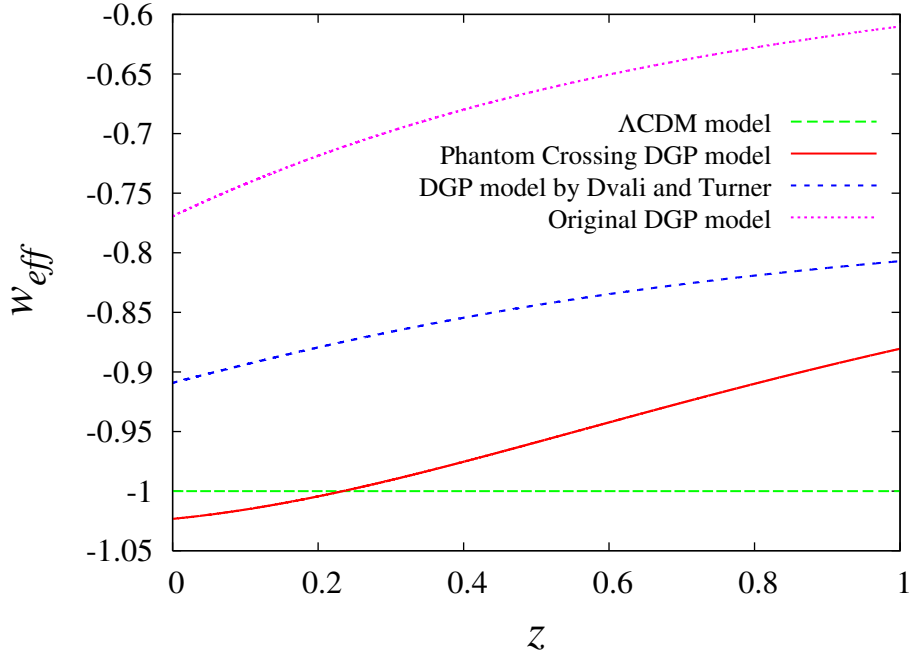}
\caption{The effective equation of state $w_{eff}$, vs. the redshift $z$. Models and parameters are same as Fig. \ref{fig:snia}.
\label{fig:weff}}
\end{figure}
\par
Fig. \ref{fig:weff} shows the effective equation of state $w_{eff}$ versus the redshift $z$. Models and parameters are same as Fig. \ref{fig:snia}. Only our Phantom Crossing DGP model can realize crossing of the phantom divide line at $z \sim 0.2$ as indicated by recent observations.
\section{Summary \label{sec:5}}
\begin{itemize}
\item We confirmed that the effective equation of state does not cross the phantom divide line in the original DGP model. \\
\item We also demonstrated that crossing of the phantom divide does not occur in the DGP model by Dvali and Turner. \\
\item We constructed the Phantom Crossing DGP model. This model realizes crossing of the phantom divide. We found that the smaller the value of the new introduced parameter $\beta$ is, the older epoch crossing of the phantom divide occurs in. Our model can realize late-time acceleration of the universe very similar to that of $\Lambda$CDM model, without dark energy, due to the effect of DGP gravity. In the proposed model (($\beta, \Omega_{m,0}$) = ($0.50, 0.30$)), crossing of the phantom divide occurs at $z \sim 0.2$ as indicated by recent observations.
\end{itemize}

\begin{acknowledgements}
The authors would like to thank the anonymous reviewer for their helpful comments and discussions.
\end{acknowledgements}

%\section*{References}

%\bibliography{hirano}   % name your BibTeX data base

\end{document}